\documentclass[useAMS,usenatbib,referee]{mn2e}
\usepackage{graphicx,graphics,aas_macros,amsmath}
\usepackage[authoryear]{natbib}

\usepackage{pdflscape}  
\usepackage{txfonts}
\usepackage{rotating}
\usepackage{times}
\usepackage{booktabs}
\usepackage{longtable}
\usepackage{color}
\usepackage{natbib}
\usepackage{amsmath}
\usepackage[utf8]{inputenc}
\usepackage{float}
\usepackage{lmodern}

\title[OAO 1657–415]{Multitude of iron lines including a Compton scattered component in OAO 1657–415 detected with \emph{Chandra}}
\author[P. Pradhan, G. Raman, B. Paul]{Pragati Pradhan$^{1,2}$ \thanks{E-mail:pup69@psu.edu;}, Gayathri Raman$^{3}$, Biswajit Paul$^{3}$ \\
$^{1}$ Penn State University, State College, Pennsylvania, 16801, US \\ 
$^{2}$ St. Joseph's College, Singamari, Darjeeling-734104, West Bengal, India\\
$^{3}$ Raman Research Institute, Sadashivnagar, Bangalore-560080, India
 }
\begin{document}
\date{}
\maketitle
\label{firstpage}
\begin{abstract}
 We present a high resolution X-ray spectrum of the accreting X-ray pulsar, OAO 1657–415 with HETG+ACIS-S onboard \emph{Chandra}, 
 revealing the presence of a broad line component around $\sim$ 6.3 keV associated with the neutral iron K$_\alpha$ line at 6.4 \rm{keV}. 
 This is interpreted as Compton shoulder arising from the Compton scattering of the 
 6.4 \rm{keV} fluorescence photons making OAO 1657–415 the second accreting neutron star where such a feature is detected. A Compton shoulder reveals 
 the presence of dense matter surrounding the X-ray source. 
 We did not detect any periodicity in the lightcurve and obtained an upper limit of $\sim$ 2\% for the pulse fraction during this observation. 
 This could be due to the smearing of the pulses when X-ray photons are scattered from a large region around the neutron star.  
 In addition to the Fe K$_\alpha$, Fe $K_\beta$ and Ni $K_\alpha$ lines already reported for this source, we also report for the first time, 
 the presence of He-like and H-like iron emission lines at 6.7 and 6.97 \rm{keV} in the first order HETG spectrum. 
 The detection of such ionized lines, indicative of a highly ionized surrounding medium, is rare in X-ray binaries.
 \end{abstract}

\section{Introduction}
The X-ray spectrum of an accreting neutron star in a High Mass X-ray Binary (HMXB) in most cases can usually be described by 
a power-law with a photon index in the range of 0.5-1.5 and a high energy cutoff at typically 10-20 \rm{keV} 
along with a Cyclotron Resonant Scattering Feature and iron K$_\alpha$ emission line at 6.4 \rm{keV}.
A few HMXBs, with low absorption column density also often show a soft excess \citep[see eg.,][]{paul2002}.
The iron emission lines produced by fluorescence emission, play a significant role in probing the
matter surrounding the neutron star. Additionally, at very large values of the ionisation parameter, we see He-like and H-like iron emission lines
at 6.7 and 6.97 \rm{keV}. Some of these ionized lines are seen in a few X-ray binaries like Cen X–3 \citep{wojdowski2004,iaria2005,naik2011} 
and Cyg X–3 \citep{paerels2000}.
If the Compton optical depth of the medium is high ($>$ 0.1; \citealt{watanabe2003}), then this emitted photon of energy 6.4 \rm{keV} 
can be further Compton scattered in the same medium and lose some energy (around 156 \rm{eV} for backscattered photons; 
\citealt{watanabe2003}). Such scattered photons produce a shoulder at a lower energy termed as the Compton Shoulder. \\
\noindent
Thanks to the unmatched spectral resolution of \emph{Chandra}, the first order Compton shoulder has earlier been observed in one HMXB, GX 301–2 
\citep{watanabe2003}. 
Another source with similar high absorption and a strong iron line as GX 301–2 is OAO 1657–415. OAO 1657–415 is an eclipsing X-ray binary discovered with the 
\emph{Copernicus} satellite \citep{P78} with the companion star being an Ofpe/WN9 type supergiant \citep{MA09} and the neutron star having a pulse period of $\sim$ 38 
\rm{s} \citep{WP79}. A study of dust-scattered halo with ASCA estimated the distance to the source as 7.1 $\pm$ 1.3 \rm{kpc} \citep{A06} complying with 
the earlier estimated distance of 6.4 $\pm$ 1.5 {kpc} obtained from the study of the infrared counterpart \citep{dist_oao}. 
The orbital period of this system is $\sim$ 10.5 \rm{days} \citep{C93} with an orbital decay rate of $\dot{P}_{orb}$ $\sim$ $(-9.74\pm0.78)\times10^{-8}$ 
\citep{J12} and the eccentricity of the orbit is $\sim$ 0.1 \citep[]{C93,bildsten1997,J12}. 
The X-ray lightcurves of OAO 1657–415 show large variation in intensity even outside the eclipse \citep{BS08}. The same authors have also demonstrated 
the presence of an unexplained (possibly temporary) dip at phase $\sim$ 0.55 in the ASM lightcurve of OAO 1657–415. 
In the broad band X-ray spectrum obtained with \emph{Beppo}-SAX and \emph{Suzaku}, a possible existence of a Cyclotron Resonance Scattering Feature (CRSF) at $\sim$ 36 \rm{keV}
was seen  with limited statistical significance indicating magnetic
field strength $3.2 (1+z)\times10^{12}$ \rm{G}, where z is the gravitational
redshift  \cite[]{O99,BS08,pradhan2014}. \\
A detailed time-resolved spectroscopy of OAO 1657–415 with \emph{Suzaku} showed a 
large variation in the absorption column density along with a very large equivalent width of Fe K$_\alpha$ line at some time intervals of a long observation 
\citep{pradhan2014}. 
This indicates the neutron star passing through a dense clump giving rise to high column density, similar to the pre-periastron passage of 
GX 301–2 \citep{islam2014_gx301}. 
Such conditions are conducive for the formation of a Compton shoulder. 
This prompted us to probe into the high resolution \emph{Chandra} spectrum of OAO 1657–415, where such a feature would be discernible. \\
\section{Observation and data analysis}
OAO 1657–415 was observed with the \emph{Chandra} observatory at the same phase (0.5-0.6) where a transient dip had earlier been noticed 
in the ASM lightcurves \citep{BS08}. The observation was carried out from 2011-05-17, 13:29:11 UT till 2011-05-18, 03:48:54 UT with 
High Energy Transmission Grating (HETG; \citealt{canizares2005}) for 50 \rm{ks}. 
HETG consists of two transmission gratings, a medium-energy grating (MEG; 0.4-5.0 \rm{keV}), 
and a high-energy grating (HEG; 0.8-10.0 \rm{keV}). The dispersed grating spectra are recorded with an array of CCDs - Advanced CCD Imaging
Spectrometer (ACIS-S; \citealt{garmire2003} ). \\
Data was reprocessed using the package \emph{Chandra} Interactive Analysis of Observations (CIAO, version 4.8) following the CXC guidelines 
\footnote{http://cxc.harvard.edu/ciao/threads/}. The evt1 files were reprocessed with \texttt{chandra\_repro} script
which automates the recommended reprocessing steps and results in processed evt2 files ready 
for scientific analysis. This observation was made in `Timed exposure' in half of the ACIS-S CCDs (512 number of rows) beginning with the first row of the CCD. \\
To extract the ACIS-S lightcurve, we used the command \texttt{dmextract} for CCD\_ID=7 in the evt2 files with the following region selection: The 
source region was chosen at the centre of the ACIS-S image with a radius of 5 arc-seconds and the background region was chosen as an annulus with inner(outer) radii of 
7(15) arc-seconds. The ACIS-S spectrum was also extracted with the same region selection and event file using the command \texttt{specextract}. \\
For the lightcurve extraction from HETG, we defined source region as a box with length(breadth) of 660(10) arc-seconds while excluding the zero order central region 
(corresponding to the centre of the ACIS-S image). 
For the background region, we excluded this central ACIS-S source region along with the HETG source region and defined a region of the same size as the HETG source region. 
The background corrected lightcurve was then extracted with the command \texttt{dmextract}. 
The lightcurves from both the instruments were barycentre corrected using the orbit ephemeris file during the observation with the help of the 
command: \texttt{axbary}.
The spectral products for HEG for different orders is automatically obtained while executing the \texttt{chandra\_repro} script. 
The same script also creates the response matrices for the respective grating orders. 
The positive and negative diffraction first order spectra and their corresponding response files, so obtained, were added using 
\texttt{combine\_grating\_spectra}\footnote{The higher order spectra are too faint for any meaningful analysis and are therefore not used.}. \\
The resultant spectrum for both instruments were grouped with a minimum of 20 counts per bin and finally spectral fitting was then carried out with \texttt{XSPEC} v 12.9.0. \\

\section{Results}
\subsection{Timing analysis}
\begin{figure*}
\centering
 \includegraphics[height=7cm,width=7cm,angle=-90]{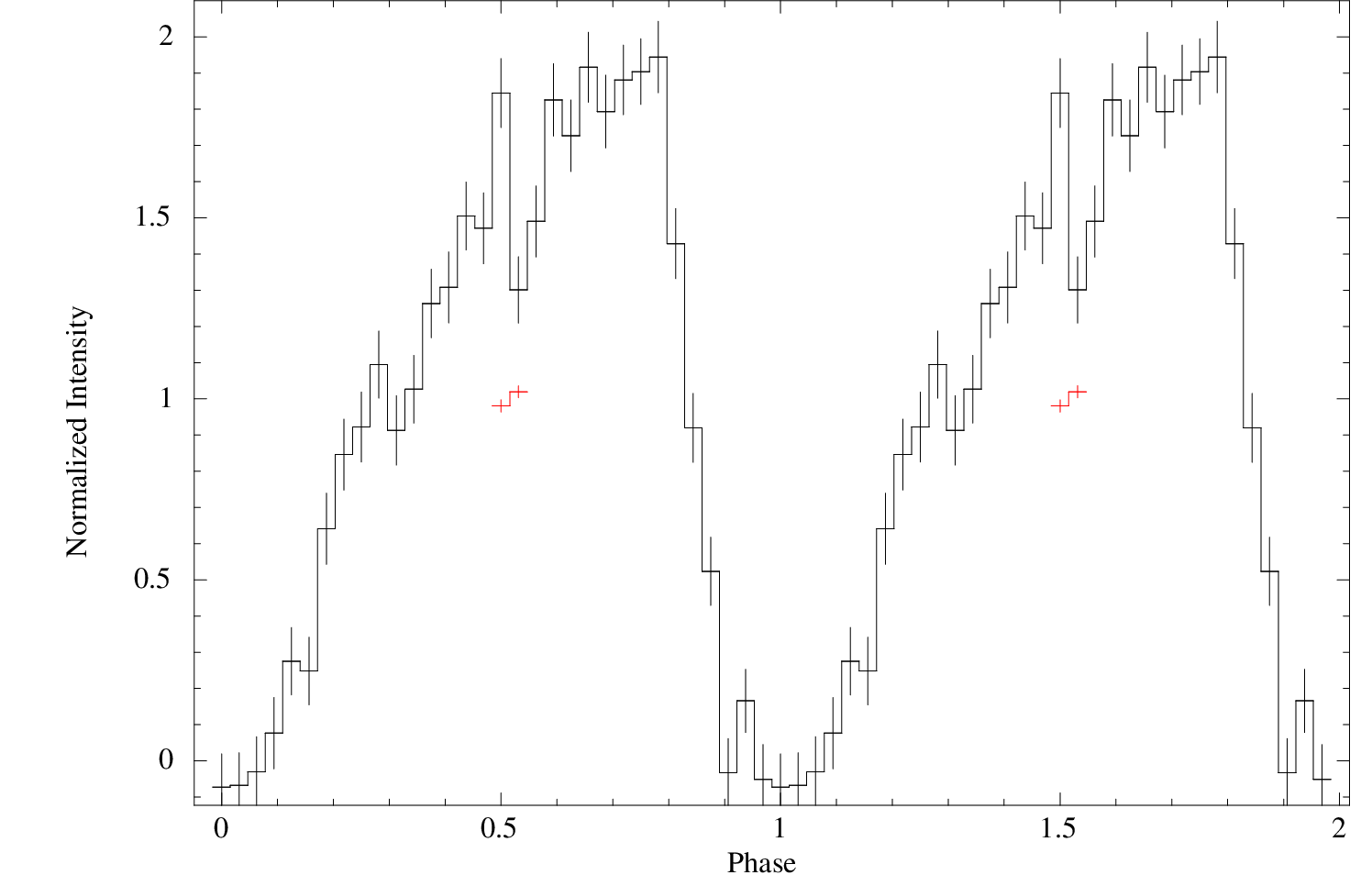}
\includegraphics[height=7cm,width=7cm,angle=-90]{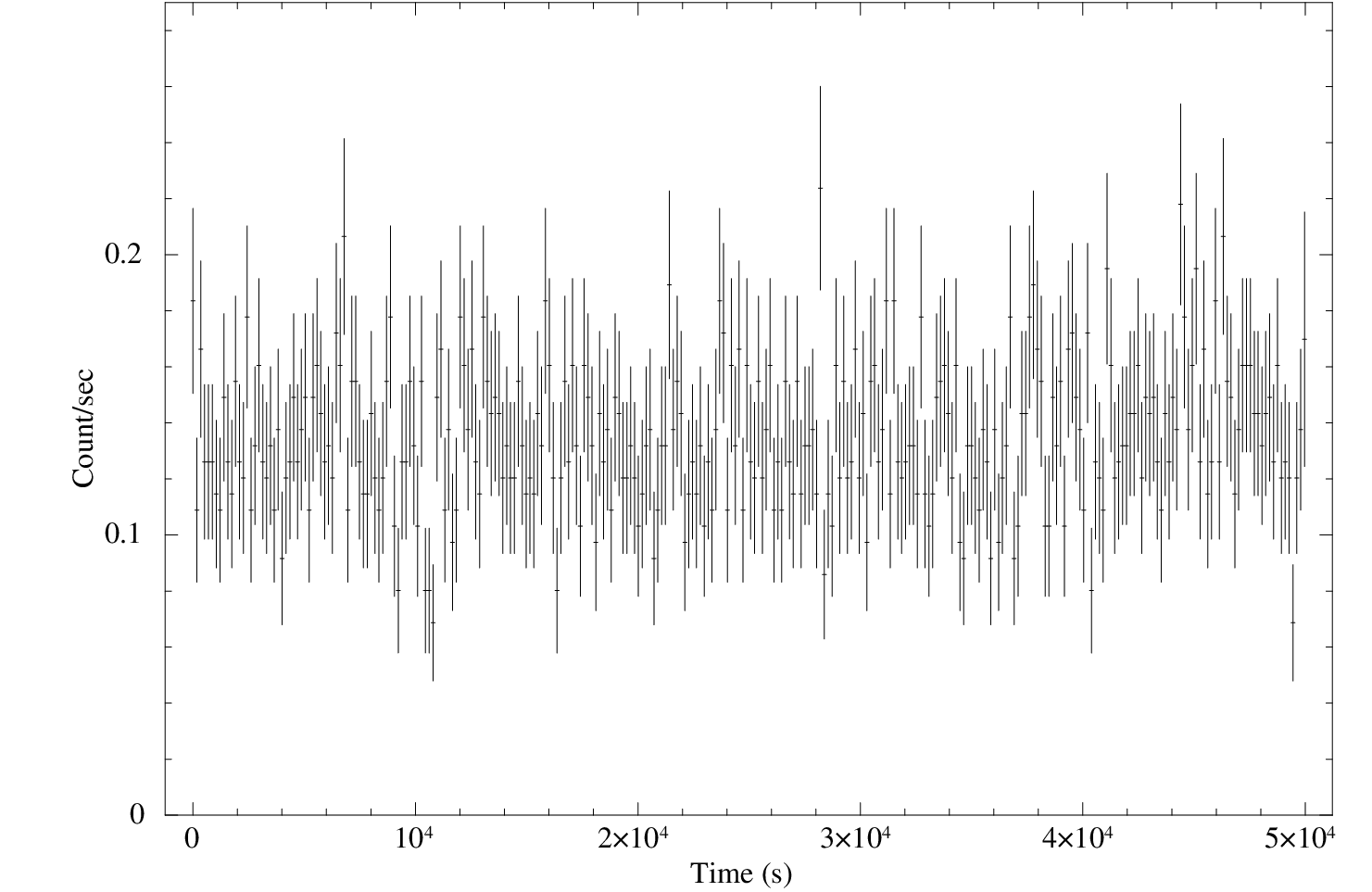}
  \caption{Left: Orbital intensity profile of OAO 1657–415 with the RXTE/ASM and \emph{Chandra}/ACIS-S lightcurves folded at an orbital period of 10.44749 d 
  \citep{falanga2015}. The phase zero corresponds to the mid-eclipse time MJD 50689.116 \citep{falanga2015}. 
  Right: ACIS-S lightcurve binned at $\sim$ 174 s. The zero of the lightcurve is $\sim$ MJD 55698.5781.}
 \label{efold}
\end{figure*}

\begin{figure*}
 \includegraphics[height=8cm,width=9cm,angle=-90]{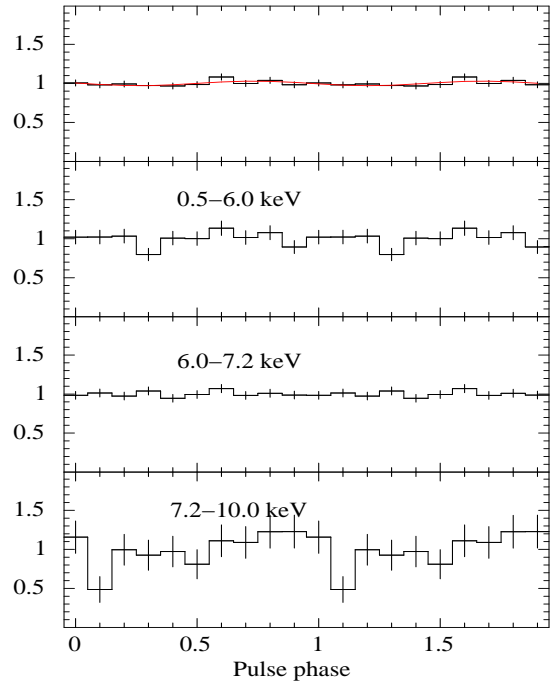}
 \caption{The uppermost panel represent the average ACIS-S lightcurve folded at a spin period of 36.968638 s estimated from the Fermi GBM history of 
 the source during this epoch. The ACIS-S lightcurve has been fit with a sine wave (plus a constant) for determining the pulse fraction. The second, third and fourth 
 panels from above are the energy resolved ACIS-S lightcurves folded at the same period. The lack of pulsations are clearly seen.}
 \label{pp}
\end{figure*}

\begin{figure*}
 \includegraphics[height=4.5cm,width=5.5cm,angle=-90]{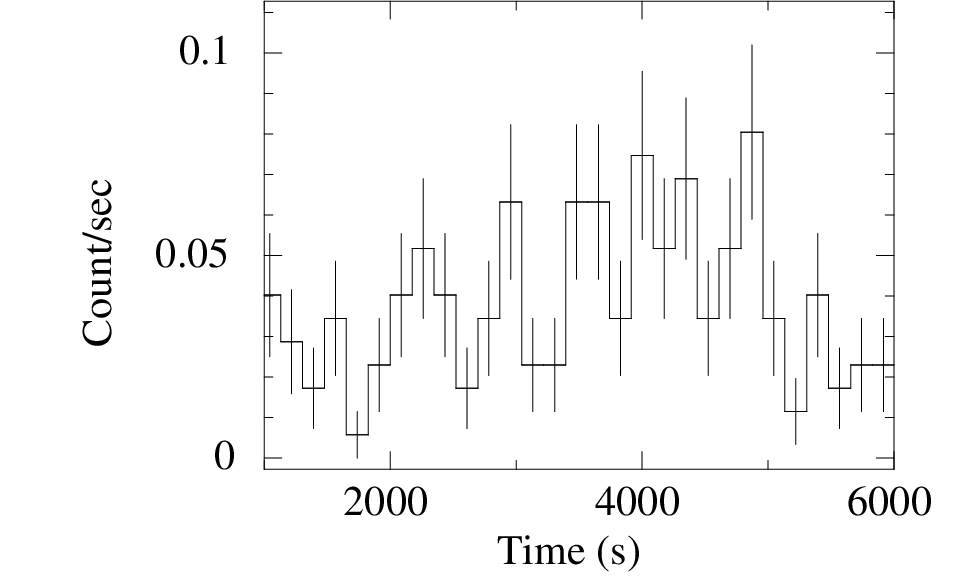}
 \includegraphics[height=4.5cm,width=5.5cm,angle=-90]{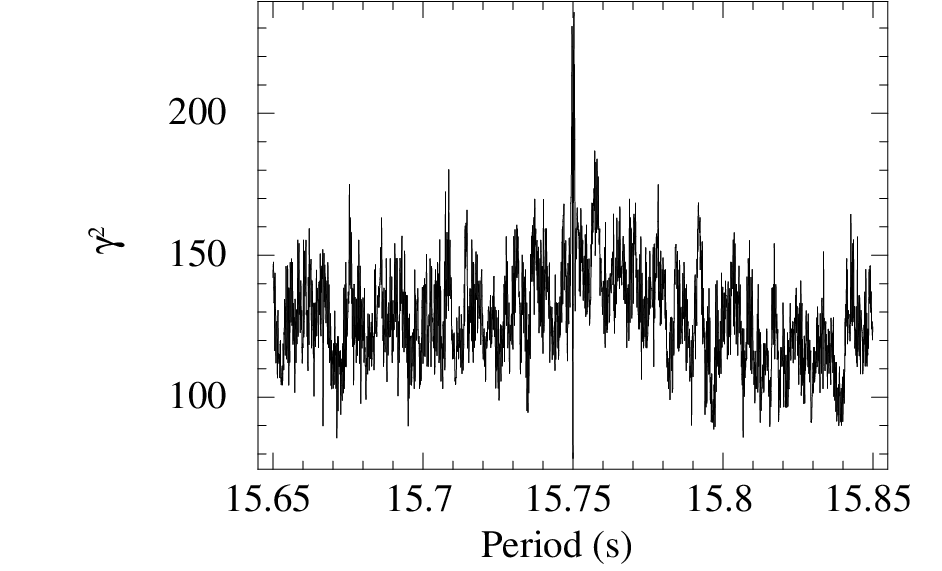}
 \includegraphics[height=4.5cm,width=5.5cm,angle=-90]{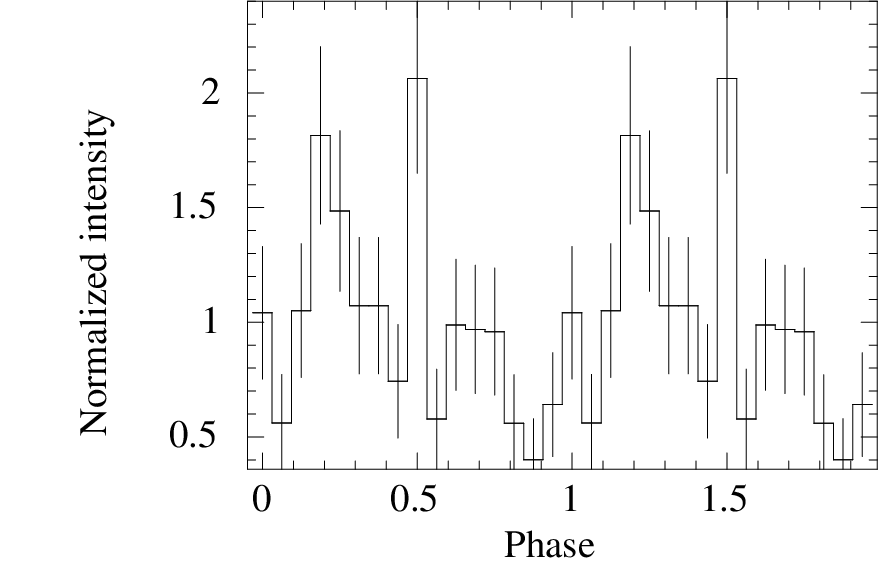}
 \caption{Left: Background subtracted lightcurve of XTE 1946+274 binned at $\sim$ 174 s, Middle: Period search with \texttt{efsearch}, Right: Folded pulse profile of XTE 1946+274. The presence of pulsations with the 
 same analysis procedure for OAO 1657–415 ensures that the lack of it in latter is not due to data reduction procedures.}
 \label{xte_lc}
\end{figure*}

We folded the ACIS-S lightcurves (right of Figure \ref{efold}) along with RXTE/ASM lightcurve at a period of 10.44749 d \citep{falanga2015} to obtain the orbital profile of 
OAO 1657–415 as is shown on the left of Figure \ref{efold}. The observation was carried out between orbital phases $\sim$ 0.5-0.6 where the 
phase zero correspond to the mid-eclipse time MJD 50689.116 \citep{falanga2015}. On the right of the same figure, we have plotted the ACIS-S lightcurve binned 
at 100 times the original binning at $\sim$ 174 s. We searched for periodicity in both the ACIS-S and HETG lightcurve (binned at 1.74 s) using FTOOLS task 
\texttt{efsearch} and did not detect any. 
We then looked into the pulse period evolution history of the source measured with the Fermi Gamma Ray Burst Monitor (GBM) and found the spin period of this pulsar 
during the time of this observation to be 36.968638 s.
The zeroth order lightcurve, which has more sensitivity and better statistics than HETG, was then folded at this period as shown in Figure \ref{pp}. The 
upper limit on pulse fraction (Pmax–Pmin/Pmax+Pmin, where Pmax and Pmin correspond to the maximum and minimum pulse amplitude respectively.) for the ACIS-S 
lightcurve was about 2\%. 
Note that we have not carried out any orbital correction when searching for periodicity. This is because the decoherence 
timescale\footnote{obtained from Eqn A9 of \citealt{deepto1997}} for this system is nearly one day which is much greater than this observation duration (T) of 
$\sim$ 50 kilo-sec, and the maximum smearing\footnote{$\delta$t$_{max}$=$\frac{\mbox{axsini(1-cos$\theta$)}}{\mbox{c}}$; where  
\mbox{$\theta$ = $\frac{\mbox{T}}{{P}_{orb}}$ $\times$ 2 $\pi$}} of the pulse phase without an 
orbital motion correction ($\delta$t$_{max}$) is estimated to be around 1.6 s, which is only a small fraction of the pulsar spin period of $\sim$ 37 s. 
Additionally, we have also created the energy resolved pulse profiles of OAO 1657–415 (Figure \ref{pp}) 
in the energy range of 0.5-6.0, 6.0-7.2 and 7.2-10.0 keV. We did not detect any pulsations in these lightcurves as well. \\ 
In order to verify this lack of pulsations in OAO 1657–415, an identical data reduction and analysis procedure was carried out on the \emph{Chandra} 
observation of a pulsar with comparable pulse period and brightness: XTE J1946+274. XTE J1946+274 is transient Be-HMXB pulsar with a spin period of nearly 15.8 s \citep{smith1998}. 
A 5 kilo-sec long archival \emph{Chandra} observation (Obs-ID 14646), which was carried out when this source was in quiescence, has been analyzed for this purpose. 
We processed the evt1 files using the CIAO script \texttt{chandra\_repro} and obtained the evt2 files. We then extracted the ACIS-S zeroth order 
light curves using the tool \texttt{dmextract}. This light curve binned at 174 s is shown on the left of Figure \ref{xte_lc}. 
The FTOOL task \texttt{efsearch} was adopted to carry out a period search in the light curve. The \texttt{efsearch} procedure folds the light curve with a 
large number of trial periods around a specified approximate period, after which a constant is fitted to the folded light curve and 
the resultant $\chi^{2}$ is obtained. 
The trial period which shows the maximum $\chi^{2}$ represents the true period. Using our light curve of XTE J1946+274, we performed a narrow period search in the 
range 15.6 s to 15.8 s with a step size of 1$\times$E${-4}$ s. We found a peak in the $\chi^{2}$, which we fitted using a Gaussian and determined the pulse period to be 
15.758 $\pm$ 0.003 s. This period was then used to fold the light curve (as shown in middle and right of Figure \ref{xte_lc}) and we obtained a pulse fraction 
(defined as Imax-Imin/Imax+Imin) of 0.67 $\pm$ 0.20.
With the identical data reduction and analysis methods described above, we were able to reproduce the results of \citep{arabaci2015}. 
This re-affirms the fact that the Chandra data of OAO 1657–415 indeed lacks pulsations, possible reasons for which, are elaborated in section \ref{discussion}.

\subsection{Spectral analysis}
\begin{figure*}
\centering
\includegraphics[height=8cm,width=8cm,angle=-90]{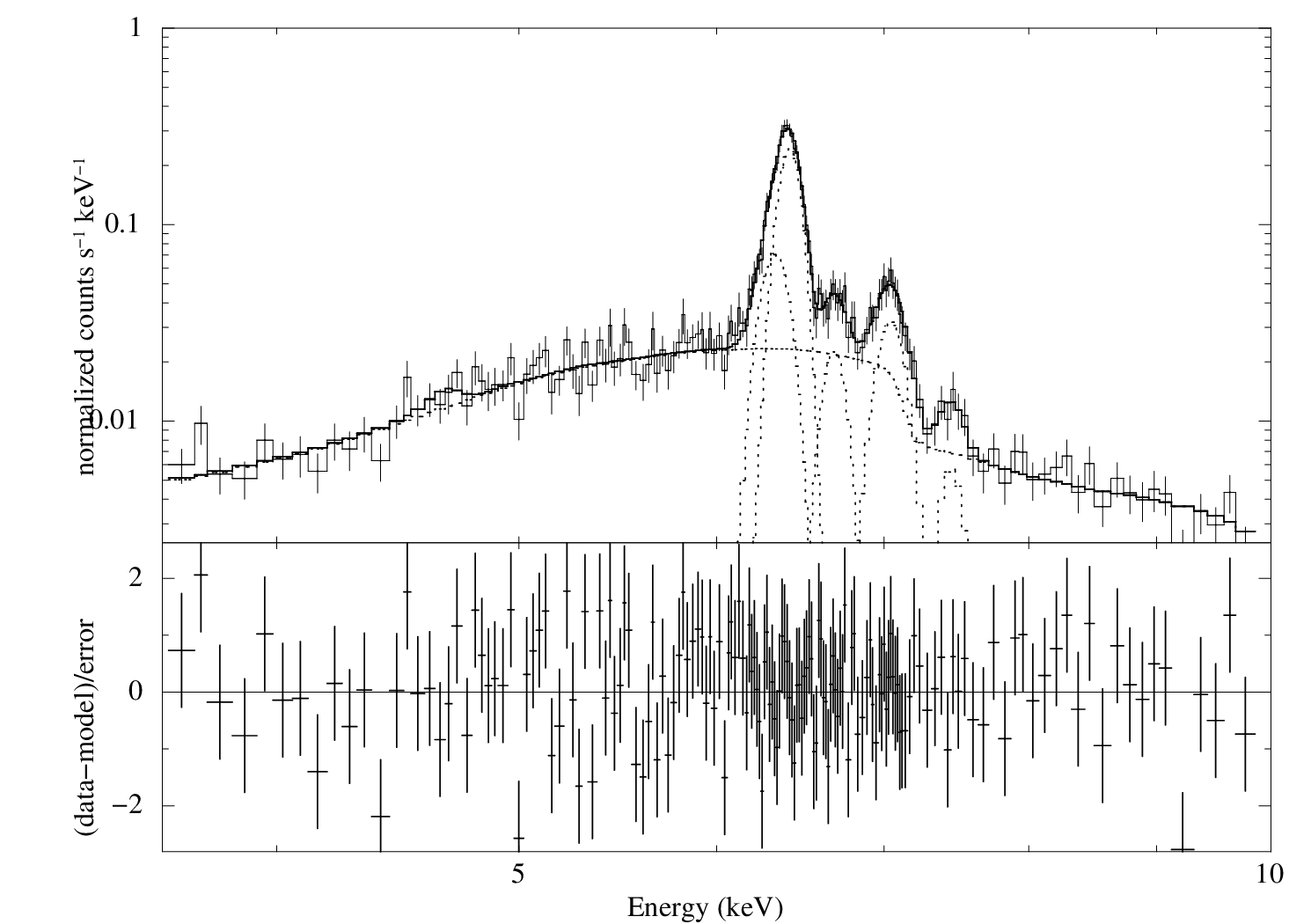}
 \includegraphics[height=8cm,width=8cm,angle=-90]{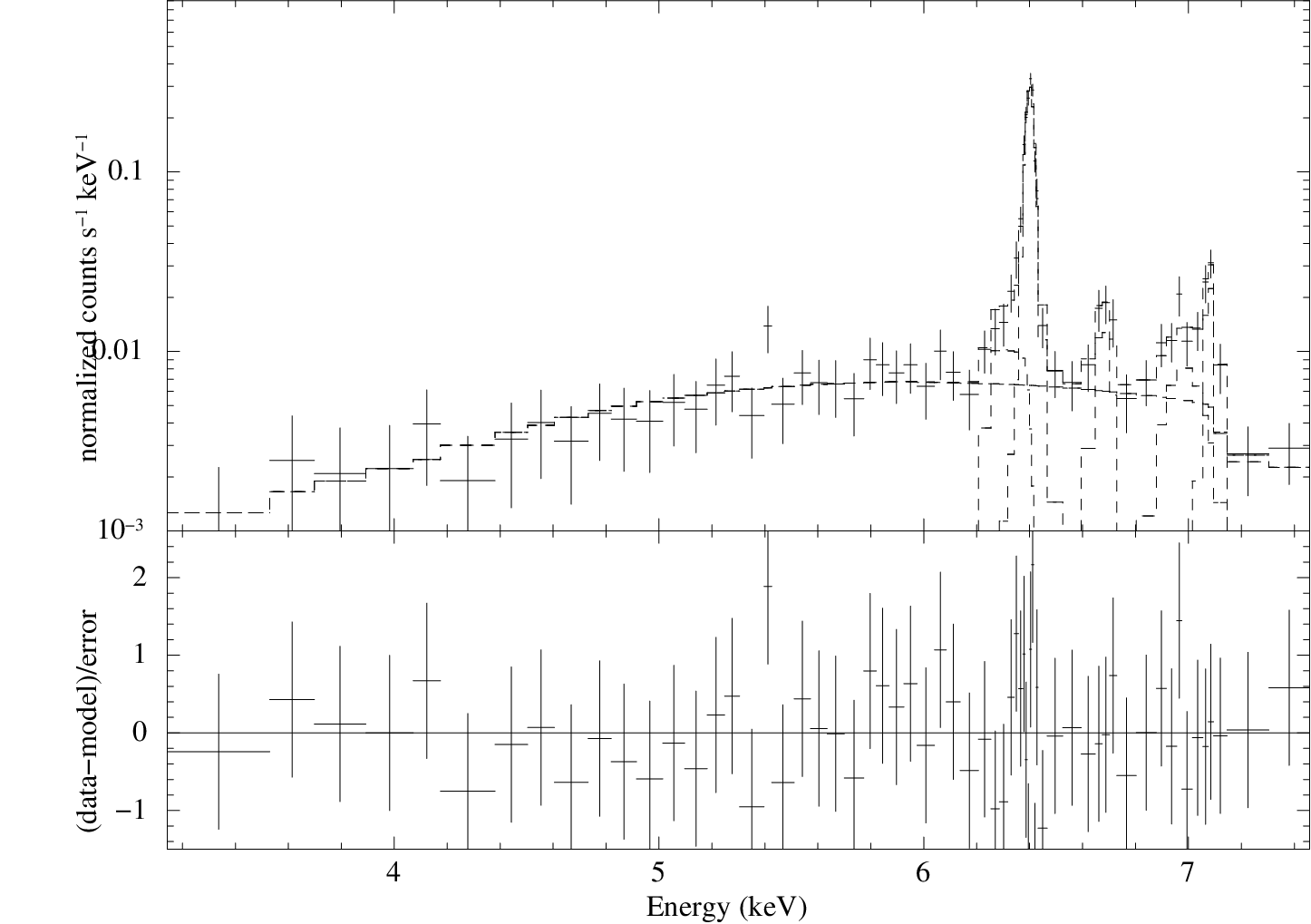}
 \caption{Left: ACIS-S spectrum of OAO 1657–415 in the energy range of 3.5-10.0 \rm{keV} with two step function that fit the Compton shoulder of the 6.4 keV line, 
 emission lines at 6.4 (Fe K$_\alpha$), 6.7 (He-like iron), 7.1 keV (Fe K$_\beta$) and 7.4 (Ni K$_\alpha$), 
 Right: HETG spectrum of OAO 1657–415 in the energy range of 3.0-7.5 \rm{keV} with with two step function that fit the Compton shoulder of the 6.4 keV line, 
 emission lines at 6.4 (Fe K$_\alpha$), 6.7 (He-like iron), 6.97 keV (H-like iron), and 7.1 keV (Fe K$_\beta$)}
 \label{acis_hetg_spec}
\end{figure*}

\begin{figure*}
 \centering
 \includegraphics[height=12cm,width=8cm,angle=-90]{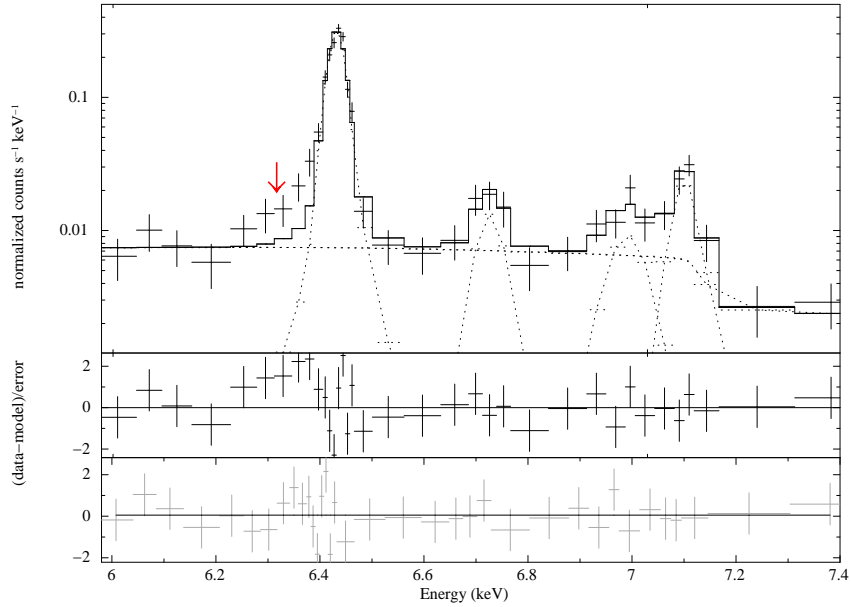}
 \caption{HETG spectrum (6-7.4 keV) of OAO 1657–415 is shown in the first panel along with the best fitted model without the two-step function for the Compton shoulder. 
 Residual to this fit is shown in the second panel and residual to the fit after inclusion of the two-step function is shown in the third panel. }
\label{doublet}
 \end{figure*}

The zero order ACIS-S and the first order HETG spectrum (HEG 1) of OAO 1657–415 were separately fit\footnote{Note that the ACIS-S (HEG 1) spectrum 
was 17 (15) \% piled up}. The X-ray spectrum is dominated by iron emission lines, with 6.0-7.2 keV X-ray photons comprising $\sim$ 58\% of the total X-ray 
photons detected in the 0.5-10.0 keV band.
To the ACIS-S continuum spectrum, we fit a powerlaw along with the line of sight photoelectric absorption and Compton scattering component (\texttt{cabs}) with solar abundances 
\emph{angr} \citep{anders1989}.
Three Gaussian lines (which are clearly seen in the raw spectrum) corresponding to K$_\alpha$ and K$_\beta$ lines of iron and K$_\alpha$ line of Nickel at 7.4 keV 
were added to obtain a reduced chisquare ($\chi^{2}_{red}$) of 2.71 while the spectrum has 153 degrees of freedom (d.o.f). 
The addition of another Gaussian line at 6.7 keV (with its width frozen to best fit value) reduced the $\chi^{2}_{red}$ (d.o.f) to 2.05 (151). 
To correct for the wavy residuals, we introduced the partial covering absorption model as is 
needed to describe the X-ray spectrum of OAO 1657–415 \citep{pradhan2014} for which the $\chi^{2}_{red}$ (d.o.f) is now 1.1 (149). 
Interestingly, we noticed that the residuals still show an asymmetry near the iron 
K$_\alpha$ line. To fit this, we added another Gaussian line centering at $\sim$ 6.3 keV 
and a $\chi^{2}_{red}$ of 0.94 was obtained with 146 d.o.f. 
On further inspection, a better fit to the same asymmetry was obtained with a two Step 
function (convolved with Gaussian) in \texttt{XSPEC} such that one of the line energies is tied to the line energy of the 
K$_\alpha$ iron line, and the other is fixed at (K$_\alpha$ -- 0.16 keV) with the widths of both as zero and their normalizations equal but opposite in sign. 
For this model, the $\chi^{2}_{red}$ obtained was 0.92 for 148 d.o.f. We choose the latter model to represent the additional broad low energy component associated 
with the iron K$_\alpha$ line. \\
The HETG spectrum was fitted independently, with the same spectral model as the ACIS-S spectrum, with the difference that since the K$_\alpha$ line of Nickel was outside 
the band used for HETG fitting, we did not detect it. Instead, we detected another line at 6.97 keV in the HETG spectrum, thanks to the better spectral resolution 
of HETG versus ACIS-S. The 
X-ray spectrum of both the instruments are shown in Figure \ref{acis_hetg_spec} and the details about the the spectral parameters - with and without using an 
additional component at $\sim$ 6.3 keV - are given in Table \ref{table:oao}. 
The analytical form of the model is: 

\hspace{-1cm}
\begin{equation}
 {\ensuremath{
   \mathrm{S(E)} = e^{-N_{\rm H1}\sigma(E)}\times(K\ E^{-\Gamma}\times fe^{-N_{\rm H2}\sigma(E)}+(1-f))+ 4~ Ga+ 2~ Step~function))
   }}
\end{equation} 
where $\mathrm{K}$, $N_{\rm H1}$, $\sigma(E)$, and $\Gamma$ are the normalization (in units of $photons~keV^{-1}~cm^{-2}~s^{-1}~at~1~\rm~{keV}$), Hydrogen 
column density (in units of $10^{22}$ atoms $cm^{-2}$), photo-electric with compton scattering cross-section, and photon index, respectively while $N_{\rm H2}$ is the 
partial covering column density (in units of $10^{22}$ atoms $cm^{-2}$) with the covering fraction as $f$. \\
The plausible explanation for the additional component at $\sim$ 6.3 keV is the presence of a Compton shoulder \footnote{We also cross-checked the presence of a Compton shoulder by taking into consideration the doublet structure of 
K$_\alpha$ and  K$_\beta$ lines of iron. For this, we fitted K$_\alpha$ (K$_\beta$) line with two Gaussian lines, the energy of one less than other 
by 13.2(16.0) eV and the relative normalisations in the ratio 2:1(2:1) as suggested in \citet{barra2009}. 
The asymmetry in the K$_\alpha$ line is still evident in the X-ray spectrum consistent 
with our fitting performed by using one Gaussian each for K$_\alpha$ and K$_\beta$ lines.}. 
The individual fits are shown in Figure \ref{acis_hetg_spec}. \\
 To demonstrate this, we remove the Compton 
shoulder line from the fit and plot the resultant spectrum in Figure \ref{doublet}. The positive excess marked with arrow is evidence of the need 
for an additional, somewhat broad component at $\sim$ 6.3 keV which is the Compton shoulder. \\
The Compton shoulder is detected in both ACIS-S and HETG spectra independently with reduction in chisq of 22 and 16 respectively for addition of one component. 
The probability of chance improvement (PCI) using F-test in \texttt{XSPEC} on addition of the Compton shoulder in the ACIS (HETG) spectrum is 
2.4E-6 (2.9E-5). This low value of PCI also adds significance to the presence of Compton shoulder.\\

\clearpage

\begin{longtable}{c|c|c|c|c}
\caption{Best fit phase averaged spectral parameters of OAO 1657–415. Errors quoted are for 90 per cent confidence range.} \\
Parameter & HETG & HETG & ACIS-S & ACIS-S   \\
& & without Compton & &without Compton  \\
& & shoulder & & shoulder\\
\hline
\vspace{-0.25cm}
$N_{\rm H1}$$^a$  & $8_{–6}^{+6}$ & $8_{–6}^{+7}$  & $10_{–4}^{+4}$ & $8_{–2}^{+3}$  \\	
\vspace{-0.1cm}
$N_{\rm H2}$$^a$  & $48_{–4}^{+5}$ & $48_{–4}^{+5}$ & $66_{–3}^{+3}$& $70_{–2}^{+2}$\\
\vspace{-0.1cm}
$f$  & $0.95_{–0.02}^{+0.05}$ & $0.95_{–0.05}^{+0.04}$& $0.96_{–0.01}^{+0.01}$& $0.96_{–0.01}^{+0.01}$\\
\vspace{-0.1cm}
$\Gamma$ & $0.21_{–0.07}^{+0.08}$ & $0.19_{–0.05}^{+0.06}$ & $0.27_{–0.14}^{+0.14}$ & $0.38_{–0.14}^{+0.14}$\\
\vspace{-0.1cm}
$\Gamma^{b}_{norm}$ & 0.0013 $\pm$ 0.0001 & 0.0011 $\pm$ 0.0001  & $0.0014_{–0.0003}^{+0.0004}$ & 0.0019 $\pm$ 0.0004 \\
Step Line flux$^{c}$ & $8.8_{–4.8}^{+1.7}$ & - & $16.3_{–1.8}^{+1.8}$ & - \\
Line centre (keV) &  6.40 $\pm$ 0.01 & 6.40 $\pm$ 0.01 & 6.41 $\pm$ 0.01 & 6.39 $\pm$ 0.01\\
\vspace{-0.1cm}
Line width (keV) & 0.008 $\pm$ 0.002& 0.009 $\pm$ 0.002  & $0.013_{–0.03}^{+0.04}$ & $0.041_{–0.005}^{+0.005}$\\
\vspace{-0.1cm}
Line norm$^{d}$ & 65.3 $\pm$ 4.5 & 68.0 $\pm$ 4.4 & $49.3_{–5.9}^{+4.3}$ & $63.8_{–2.5}^{+2.6}$\\
\vspace{-0.1cm}
EW (keV) & 1.12 $\pm$ 0.07 &1.46 $\pm$ 0.08 & $0.79_{–0.09}^{+0.07}$& $1.73_{–0.05}^{+0.05}$\\
\vspace{-0.1cm}

Line centre (keV) & $6.68_{–0.01}^{+0.02}$ & $6.68_{–0.02}^{+0.02}$ & $6.69_{–0.02}^{+0.02}$ & $6.69_{–0.02}^{+0.02}$ \\
\vspace{-0.1cm}
Line width (keV) & 0.022 & 0.022 $\pm$ 0.015 & 0.025 & 0.001\\
\vspace{-0.1cm}
Line norm$^{d}$ & 6.6 $\pm$ 2.8 & 6.1 $\pm$ 2.7 & $5.8_{–1.1}^{+1.1}$ & $4.9_{–1.1}^{+1.2}$ \\
\vspace{-0.1cm}
EW (keV) & 0.04 $\pm$ 0.02 & 0.04 $\pm$ 0.01 & 0.05 $\pm$ 0.01 & 0.04 $\pm$ 0.01 \\
\vspace{-0.1cm}

Line centre (keV) & 6.97 $\pm$ 0.03 & 6.97 $\pm$ 0.03 & - & -\\
\vspace{-0.1cm}
Line width (keV) & $0.068_{–0.023}^{+0.045}$ & $0.062_{–0.037}^{+0.044}$ & - & - \\
\vspace{-0.1cm}
Line norm$^{d}$ & $11.3_{–5.5}^{+4.4}$ & $10.2_{–4.6}^{+5.1}$ & - & -\\
\vspace{-0.1cm}
EW (keV) & 0.21 $\pm$ 0.08 & 0.18 $\pm$ 0.09  & -& -\\
\vspace{-0.1cm}

Line centre (keV) & 7.08 $\pm$ 0.01 & 7.07 $\pm$ 0.01 & $7.05_{–0.01}^{+0.01}$ & $7.06_{–0.01}^{+0.01}$\\
\vspace{-0.1cm}
Line width (keV) & 0.0003 & 0.0003 &$0.06_{–0.02}^{+0.02}$ & $0.06_{–0.02}^{+0.02}$\\
\vspace{-0.1cm}
Line norm$^{d}$ & 9 $\pm$ 3 & 9.3 $\pm$ 3.4 & 14.2 $\pm$ 1.8 & 14.6 $\pm$ 2.1\\
\vspace{-0.1cm}
EW (keV) & 0.18 $\pm$ 0.07 & 0.20 $\pm$ 0.07 & 0.46 $\pm$ 0.06 & 0.46 $\pm$ 0.06\\
\vspace{-0.1cm}

Line centre (keV) & - & - & $7.45 \pm 0.04$ & 7.45 $\pm$ 0.4\\
\vspace{-0.1cm}
Line width (keV) & - & -&0.04 & 0.04\\
\vspace{-0.1cm}
Line norm$^{d}$ & - & -&3.0 $\pm$ 1.5 & 3.1 $\pm$ 1.2\\
\vspace{-0.1cm}
EW (keV) & - & -& $0.12 \pm$ 0.04 & 0.12 $\pm$ 0.04\\

\vspace{-0.1cm}
$\chi^{2}_{\nu}$/d.o.f & 0.74/47 & 1.1/48 & 0.92/148 & 1.1/149\\
\vspace{-0.1cm}
Flux$^{e}$ (1-10 keV) & 2.99 $\pm$ 0.12 & 2.99 $\pm$ 0.24 & 2.41 $\pm$ 0.05 & 2.42 $\pm$ 0.05 \\
\bottomrule
\label{table:oao}
\end{longtable}
$^a$ In units of $10^{22}$ atoms $cm^{-2}$ \\
$^b$ \ensuremath{\mathrm{photons}\, \mathrm{keV}^{-1}\,\mathrm{cm}^{-2}\,\mathrm{s}^{-1}\,\mathrm{at}\, 1\, \mathrm{keV}} \\
$^c$ \ensuremath{\mathrm{photons}\, \mathrm{cm}^{-2}\, \mathrm{s}^{-1}\ \times10^{-5}} with step line energy at 6.4 and (6.4-0.16) for the width of both fixed at zero.  \\
$^d$ \ensuremath{\mathrm{photons}\, \mathrm{cm}^{-2}\, \mathrm{s}^{-1}\ \times10^{-5}}  \\
$^e$ In units of \ensuremath{10^{-11}\,  \mathrm{erg}\,  \mathrm{cm}^{-2}\,\mathrm{s}^{-1}} \\

\section{Discussion}
\label{discussion}
Supergiant stars have strong stellar winds which are often inhomogenous and when the neutron star passes through clumps of varying sizes and wind density, it leads to variable accretion, 
variable column density and emission line strength. If the neutron star passes through a very dense clump of matter, reprocessing of the source photons 
in the material will produce iron fluorescence line and a Compton shoulder will be produced if a large enough fraction of the K$_\alpha$ 
iron line photons are scattered by electrons in the dense medium surrounding the X-ray source.
A careful analysis of the ACIS-S and HETG spectrum of OAO 1657–415 has led to the detection of this Compton shoulder along with a very high column density of absorption. 
This observation is carried out at $\sim$ phase 0.55 in the ASM lightcurves (left of Figure \ref{efold}) at which phase, 
there are reports of temporary dips \citep{BS08}. Timing analysis of OAO 1657–415 allows us to place an upper limit on the pulse fraction ($\sim$ 2 \%; Figure \ref{pp}). 
The low pulse fraction is reminescent of one part of a long Suzaku observation (segment C in \citealt{pradhan2014}) interpreted as the source passing through a dense clump of 
matter where the pulse fraction was the lowest. During the same segment C, the equivalent width of the K$_\alpha$ iron line was also the strongest which implies that the neutron star is passing \emph{through} a clump (see 
Section 4.1 of \citealt{pradhan2014} for details). These highly absorbed states seem to be present at various phases (segment C is carried out around phase 0.25, 
and the current observation is at phase $\sim$ 0.55) indicating that the orbit of OAO 1657–415 have clumps of matter at different phases.
In a luminous HMXB system, the wind near the neutron star is photoionized by X-rays from the neutron star \citep[see, eg.,][]{watanabe2006}. 
During this \emph{Chandra} observation, in addition to the K$_\alpha$ and K$_\beta$ iron lines, 
we have also detected K$_\alpha$ emission from highly ionized ions: H-like Fe corresponding to 6.97 \rm{keV}. Among the three He triplets at 6.63 \rm{keV}, 
6.67 \rm{keV} and 6.70 \rm {keV}, we were able to resolve the emission line at 6.70 \rm {keV}. Apart from this, K$_\alpha$ lines 
of Nickel are also seen.  \\
So far, a Compton shoulder has been reported in the X-ray specrum of only one X-ray binary GX 301–2, as an asymmetric K$_\alpha$ iron line \citep{watanabe2003}. 
A detailed spectral analysis of GX 301–2 indicated that the shape of the Compton shoulder depends largely on the 
absorption column density (which contributes to the number of scatterings) and consequent smearing with the increase in the electron temperature. Several studies 
have been carried out on the dependence of the Compton shoulder on spatial and temporal parameters by assuming different geometries for the reflectors through 
detailed Monte-Carlo simulations \citep[see eg.,][]{odaka2011,odaka2016}. Such simulations 
can be best compared with X-ray spectra acquired with a very high spectral resolution, like those obtained with micro-calorimeters and broadband spectroscopy (to assume a proper spectral slope 
for the illuminating spectrum). Using the current Chandra data for OAO 1657–415, which is comparatively dimmer (count rate $\sim$ 1$/$10 of GX 301–2, 
when compared to the lightcurves obtained with \citealt{watanabe2003}), we have limited scope to perform such detailed simulations and therefore leave such estimates to 
further studies. \\
In the subsequent paragraphs, we will discuss the physical interpretations arising from the presence of the iron emission lines.

\subsection{Relative fluxes of emission lines}
\label{discussion1}
\begin{itemize}

\item The flux of the Compton shoulder is about 25\% (14\% and 33\% respectively in the HETG and ACIS-S spectrum)
of the K$_\alpha$ keV line 

\item K$_\beta$ and K$_\alpha$ of Fe: 
The ratio of K$_\beta$ to K$_\alpha$ line fluxes ($\eta$) is about 0.20 $\pm$ 0.03 (0.13 $\pm$ 0.04 and 0.28 $\pm$ 0.02 respectively in the HETG and 
ACIS-S spectrum). The expected value of this flux ratio (theoretically) for neutral gas Fe atoms is $\eta$ = 0.125 \citep{kaastra1993} while 
experimental results for solid Fe demonstrate $\eta$ = 0.1307 \citep{pawl2002}. 
 
\item Fe XXVI (6.97 keV) and Fe XXV (6.7 keV) of Fe: 
For iron to produce flourescent lines at 6.4 \rm{keV}, the ionisation parameter is required to be, $\xi (= \frac{L} {nr^2}$) $\leqslant 10^{2}$ \citep{kallman1982}. 
In the present case, the ratio of the flux of Fe XXVI (6.97 keV) to Fe XXV (6.7 keV) $\sim$ $1.76_{+0.12}^{-0.01}$ (from HETG spectrum), which correspond to $\xi \ge 10^{3.5}$ (refer to 
Figure 8 of \citealt{ebisawa1996}). Such a high degree of ionization parameter as well as the presence of Fe XXV and Fe XXVI suggest 
that the 6.4 keV line iron lines in OAO 1657–415 are produced in a region significantly farther away while the the He-like and H-like lines are produced 
from a different region, nearer to the NS. 

\item K$_\alpha$ of Ni and Fe: The ratio of intensities of K$_\alpha$ of Ni to Fe (from ACIS-S data) is $\sim$ 0.049 $\pm$ 0.014 consistent 
with solar abundances \citep{molendi2003}. \\
Lastly, the possible iron K-absorption edge at $\sim$ 7.1 \rm{keV} seen in the X-ray spectrum is consistent with the large absorption column densities. \\
\end{itemize}
\subsection{Size of the 6.4 keV line region}
In addition to the highly ionized lines of Fe as discussed earlier, the neutral K$_\alpha$ line of iron is also very strong with an equivalent width of more than 1 keV.
An approximate calculation as follows also give us an estimate of the minimum size of the clump and the estimation of the radius where the iron lines of 
varying ionization are formed. \\
We assume that the very high absorption column density and the line equivalent width is caused as the neutron star is passing through a dense clump. 
The hardness ratio does not change significantly during the observation (of $\sim$ 50 kilo-sec) making the minimum size of the clump (d) as $\sim$ 
5$\times10^{12}$ cm (assuming a nominal characteristic velocity of the stellar wind as 1000 km s$^{-1}$; \citealt{bozzo2016}). 
Assuming the clump to be spherical and the neutron star passing through its centre, the absorption is caused by matter along the radius (d/2),  
the maximum number density (n) of the clump can be estimated as $N_{\rm H}$/0.5$\times$d (0.7$\times10^{24}$/2.5$\times$10$^{12}$) $\sim$ 2.8$\times10^{11}$ cm$^{-3}$. \\
Therefore, for an absorbed luminosity (1-10 keV) $L$ of $\sim$ 1.5$\times10^{35}$ and required ionization parameter $\xi = \frac{L} {nr^2}$ $\leqslant 10^{2}$ 
the limiting inner radius for the 6.4 keV line producing region is $\sim$ 2.5 lt-sec. 
The 6.4 keV line is produced from a region outside this radius while the H-like and He-like lines are produced from a region very close to the neutron star. 
One other HMXB in which very distinct regions for different iron emission lines have been ascertained is Cen X–3 \citep{naik2011}. 
The H-like and He-like iron emission lines in Cen X–3 are produced in a region larger in size compared to the size of the companion star, 
while the K$_\alpha$ line is produced closer to the neutron star, perhaps in the outer accretion disk \citep{iaria2005}. \\
For OAO 1657-415, we should note that since the region below 2.5 lt-sec is dominated by the ionized H-like and He-like lines, we could in principle also 
expect Compton shoulder for this lines. In this case, however, these 6.7 keV and the 6.97 keV lines are 
too weak to detect the same in the \emph{Chandra} spectrum. \\

\section{Conclusion}
Through a detailed spectral fitting of the ACIS-S and first order HETG spectrum, we report, for the first time, the detection of a Compton shoulder 
in OAO 1657–415 making it the only second X-ray binary in which this feature 
is detected. We report a large absorption column density in the X-ray spectrum and the iron 
K$_\alpha$ line has a large equivalent width of $\sim$ 1 keV. In addition, we also report the detection of He-like and H-like lines of iron at 6.7 and 6.97 keV for 
the first time for this source. 
Through timing analysis, we report non-detection of pulsations (with pulse fraction less than 2\%), perhaps due to the neutron star being in a dense environment 
leading to many scatterings and subsequent loss of coherence.
Finally, we also put a lower limit on the emission line region of the iron K$_\alpha$ line at a distance greater than 2.5 lt-sec from the neutron star.

\section{Acknowledgements}
The authors would like to thank the reviewer for his contributions to the paper in the form of useful comments and suggestions. 
This research has made use of data and software provided by the High Energy Astrophysics Science Archive Research Center (HEASARC), which is a service of the 
Astrophysics Science Division at NASA/GSFC and the High Energy Astrophysics Division of the Smithsonian Astrophysical Observatory.
This research has also made use of data obtained from the Chandra Data Archive and the software provided by the Chandra X-ray Center (CXC) 
in the application packages CIAO. PP would like to thank Tanmoy Chattopadhyay for useful discussions during the preparation of the manuscript.

\bibliography{ref.bib}{}
\bibliographystyle{mn2e}

\end{document}